\documentclass[10pt]{article}

\usepackage{amsmath}
\usepackage{array}
\usepackage{appendix}
\usepackage{graphicx}
\usepackage{amsfonts}
\usepackage{amssymb}
\usepackage{mathrsfs}
\usepackage{yfonts}
\usepackage{euscript}
\usepackage{upgreek}
\usepackage{slantsc}
\usepackage{calligra}
\usepackage[T1]{fontenc}
\usepackage{epsf}
\usepackage{latexsym}

\usepackage{tipa}

\def\be{\begin{equation}}
\def\ee{\end{equation}}
\def\beq{\begin{equation}}
\def\eeq{\end{equation}}
\def\bea{\begin{eqnarray}}
\def\eea{\end{eqnarray}}

\def\!{\hspace{-1.6667em}}

\def\mC{\mbox{C}}   
\def\mD{\mbox{D}}

\def\mN{\mbox{N}} 

\def\mP{\mbox{P}}

\def\mS{\mbox{S}}

\def\me{\mbox{e}}

\def\mh{\mbox{h}}

\def\ml{\mbox{l}}   
   
\def\mn{\mbox{n}}   
\def\mo{\mbox{o}}

\def\brho{\mbox{\boldmath$\rho$}}          
\def\bh{\underline{\underline{\mbox{h}}}  }            

\def\bn{\mbox{\bf n}}

\def\bM{\mbox{\bf M}}

\def\bM{\mbox{{\bf M}}}
\def\bM{\mbox{{\bf M}}}

\def\bh{\mbox{{\bf h}}}

\def\bn{\mbox{{\bf n}}}

\def\FrQ{\mbox{\Large $\mathfrak{q}$}}

\def\FrM{\mbox{\Large $\mathfrak{m}$}}                          
\def\sFG{\mbox{$\mathfrak{g}$}}

\def\FrG{\mbox{\Large $\mathfrak{g}$}}

\def\sa{\mbox{\scriptsize a}}

\def\scc{\mbox{\scriptsize c}}

\def\se{\mbox{\scriptsize e}}
\def\sf{\mbox{\scriptsize f}}
 
\def\sh{\mbox{\scriptsize h}} 
\def\si{\mbox{\scriptsize i}}

\def\sll{\mbox{\scriptsize l}}  
\def\sm{\mbox{\scriptsize m}}
\def\sn{\mbox{\scriptsize n}} 
\def\so{\mbox{\scriptsize o}}

\def\sr{\mbox{\scriptsize r}}
\def\sss{\mbox{\scriptsize s}}  
\def\st{\mbox{\scriptsize t}}

\def\sA{\mbox{\scriptsize A}} 
\def\sB{\mbox{\scriptsize B}}

\def\sD{\mbox{\scriptsize D}}

\def\sG{\mbox{\scriptsize G}}

\def\sJ{\mbox{\scriptsize J}}
\def\sK{\mbox{\scriptsize K}}
 
\def\sM{\mbox{\scriptsize M}} 
\def\sN{\mbox{\scriptsize N}} 

\def\sP{\mbox{\scriptsize P}} 
 
\def\sR{\mbox{\scriptsize R}}
\def\sS{\mbox{\scriptsize S}}

\def\sU{\mbox{\scriptsize U}}

\def\sW{\mbox{\scriptsize W}}

\def\sfC{\mbox{\sffamily{\scriptsize C}}}

\def\sfF{\mbox{\sffamily{\scriptsize F}}}

\def\sfM{\mbox{\sffamily{\scriptsize M}}}

\def\sfO{\mbox{\sffamily{\scriptsize O}}}

\def\sfQ{\mbox{\sffamily{\scriptsize Q}}}

\def\sbM{\mbox{{\bf \scriptsize M}}}

\def\ta{\mbox{\tiny a}}

\def\te{\mbox{\tiny e}}

\def\ti{\mbox{\tiny i}}

\def\tl{\mbox{\tiny l}}

\def\tn{\mbox{\tiny n}}
\def\to{\mbox{\tiny o}}

\def\tr{\mbox{\tiny r}}

\def\ttt{\mbox{\tiny t}}   

\def\biP{\mbox{\boldmath$P$}}

\def\lft{\mbox{\Large \sffamily t}}
\def\lt{\mbox{\Large $t$}}

\def\K{Kucha\v{r} }

\def\NSI{Na\"{\i}ve Schr\"{o}dinger Interpretation }
\def\CPI{Conditional Probabilities Interpretation }

\def\pa{\partial}
\def\d{\textrm{d}}

\def\5Star{\mbox{\Large$\star$}}              
\def\cr{\mbox{\scriptsize{\bf $\mbox{ } \times \mbox{ }$}}}

\def\sumi3{\sum\mbox{}_{\mbox{}_{\mbox{\scriptsize $i$=1}}}^3}

\def\sumj3{\sum\mbox{}_{\mbox{}_{\mbox{\scriptsize $j$=1}}}^3}
\def\sumk3{\sum\mbox{}_{\mbox{}_{\mbox{\scriptsize $k$=1}}}^3}

\begin{document}

\begin{titlepage}

\begin{center}

{\Large\bf          PROBLEM OF TIME: FACETS AND MACHIAN STRATEGY}

\vspace{.1in}

{\bf Edward Anderson}

\vspace{.1in}

{\em DAMTP Cambridge}, ea212 *at* cam.ac.uk

\end{center}

\begin{abstract}

The Problem of Time is that `time' in each of ordinary quantum theory and general relativity are mutually incompatible notions. 
This causes difficulties in trying to put these two theories together to form a theory of Quantum Gravity. 
The Problem of Time has 8 facets in canonical approaches. 
I clarify that all but one of these facets already occur at the classical level, and reconceptualize and re-name some of these facets as follows.
The Frozen Formalism Problem becomes Temporal        Relationalism, 
the Thin Sandwich Problem    becomes Configurational Relationalism, via the notion of Best Matching.
The Problem of Observables becomes the Problem of Beables, and 
the Functional Evolution Problem becomes the Constraint Closure Problem.  
I also outline how each of the Global and Multiple-Choice Problems of Time have their own plurality of facets.

This article additionally contains a local resolution to the Problem of Time at the conceptual level 
and which is actually realizable for the relational triangle and minisuperspace models.
This resolution is, moreover, Machian, and has three levels: classical, semiclassical and a combined semiclassical--histories--timeless records scheme.
I end by delineating the current frontiers of this program toward resolution of the Problem of Time in the cases of full GR and of slightly inhomogeneous cosmology.  

\end{abstract}

\noindent Invited Seminar at ``Do we Need a Physics of Passage? Conference, Cape Town (December 2012).

\end{titlepage}

\section{Introduction}\label{Intro2}

This Article mostly concerns General Relativity (GR) in split 3 + 1 form. 
The Arnowitt--Deser--Misner (ADM) \cite{ADM} split metric for this is\footnote{I use lower-case Latin and Greek indices for space and spacetime objects respectively.
I also use underlines for spatial objects and bold font for configuration space objects.
$x^i$ are spatial coordinates and $X^{\mu}$ are spacetime coordinates.
I use round brackets for functions, square brackets for functionals and ( ; ] for mixed function dependence before the semicolon and functional dependence after it.} 
\beq
g_{\mu\nu}(X^{\rho}) =
\left(
\stackrel{    \mbox{$ \beta_{k}\beta^{k} - \alpha^2$}    }{ \mbox{ }  \mbox{ }  \beta_{j}    } \stackrel{    \mbox{$\beta_{i}$}    }{  \mbox{ } \mbox{ }  h_{ij}    }
\right)
\mbox{ } 
\eeq
for (see Fig 1.a) $\alpha(\underline{x}, t)$      the {\it lapse} (`time elapsed'), 
                  $\beta^{\mu}(\underline{x}, t)$ the {\it shift} (displacement in identification of the spatial coordinates between 2 adjacent slices)  
              and $h_{ij}(\underline{x}, t)$      the {\it induced metric} on the spatial hypersurface, $\Sigma$.
In this article, $\Sigma$ is taken to be of a fixed spatial topology that is compact without boundary.
The ADM split action is then 
\beq
S^{\sG\sR}_{\sA\sD\sM} := \int\d t\int_{\Sigma}\d^3x\sqrt{h}\,\alpha L^{\sG\sR}_{\sA\sD\sM} :=  \int\d t\int_{\Sigma}\d^3x\sqrt{h}\,\alpha 
\left\{
T^{\sG\sR}_{\sA\sD\sM}/\alpha^2 +  R - 2\Lambda
\right\} \mbox{ } , \mbox{ } 
\label{ADM-L}
\eeq
\beq
T^{\sG\sR}_{\sA\sD\sM} := ||\dot{\mbox{\boldmath $h$}}     - \pounds_{\underline{\beta}}\mbox{\boldmath $h$}||^2_{\mbox{\scriptsize \mbox{\boldmath $M$}}}/4 \mbox{ } .  
\eeq
Here, $\mbox{\boldmath $M$}$ has components         $M^{ijkl}(\mbox{\boldmath $h$}) := \sqrt{h}\{h_{ik}h_{jl} - h_{ij}h_{kl}\}$,    and is the GR configuration space metric -- 
the inverse of the DeWitt supermetric \cite{DeWitt} $N_{ijkl}(\mbox{\boldmath $h$}) := \{h_{ik}h_{jl} - h_{ij}h_{kl}/2\}/\sqrt{h}$. 
Also, $\dot{\mbox{ }} := \pa/\pa t$, 
      $\pounds_{\underline{\beta}}$            is the Lie derivative with respect to $\beta^i$, 
      $R(\underline{x}; \mbox{\boldmath $h$}]$ is the spatial Ricci scalar 
  and $\Lambda$                                is the cosmological constant.

The GR momenta are then 
\beq
\pi^{ij}(\underline{x}, t) := {\delta L^{\sG\sR}_{\sA\sD\sM}}/{\delta \dot{h}_{ij}} =    M^{ijkl} \{\dot{h}_{ij}  - \pounds_{\underline{\beta}} h_{ij}\}/2\alpha 
                                                                                    = - \sqrt{h}\{  K^{ij} -  K h^{ij} \}                                            \mbox{ } , 
\label{Gdyn-momenta}
\eeq
where 
\beq
K_{ij}(\underline{x}, t) := \{ \dot{h}_{ij} - \pounds_{\underline{\beta}} h_{ij}\}/{2\alpha} 
\label{K-def}
\eeq
is the extrinsic curvature of the hypersurface with metric $h_{ij}$.
GR then has a linear momentum constraint 
\beq
{\cal M}_i(\underline{x}, t; \mbox{\boldmath $h$}, \mbox{\boldmath $\pi$}]  := -2 D_j \pi^j\mbox{}_i = 0
\label{Mi}
\eeq 
from variation with respect to $\beta^i$, and a quadratic Hamiltonian constraint 
\beq
{\cal H}(\underline{x}, t; \mbox{\boldmath $h$}, \mbox{\boldmath $\pi$}] := N_{ijkl}\pi^{ij}\pi^{kl} - \sqrt{h}\{  R - 2\Lambda\} = 0 \mbox{ } .
\label{H}
\eeq
These are {\it first-class} constraints: their Poisson brackets close without producing any further constraints or other conditions.  

\mbox{ } 

\noindent The Isham--Kucha\v{r} \cite{Kuchar92I93} status quo from the 1990's of the Problem of Time (PoT, see also \cite{APoT} for a summary and update) is then as follows.
The Problem of Time has 8 facets in canonical approaches.
These are jointly underlied by the conceptual-level mismatch between time in GR and in ordinary Quantum Theory. 

\mbox{ } 

\noindent {\bf Frozen Formalism Problem}.         
GR's quadratic Hamiltonian constraint ${\cal H}$ leads to the quantum-level {\it Wheeler--DeWitt equation} \cite{DeWitt, Battelle}
\be
\widehat{\cal H}\Psi = 0
\ee
-- a subcase of time-independent Schr\"{o}dinger equation (TISE) $\widehat{H}\Psi = E\Psi$.
In other words, it is a stationary or frozen equation.  
Moreover, this occurs in a situation in which one might expect a time-dependent Schr\"{o}dinger equation $\widehat{H}\Psi = i\hbar\pa\Psi/\pa t$ for some notion of time $t$.
Attempted resolutions of this are on pages 2 and 3. 
%

\mbox{ } 
												  												  
\noindent {\bf Thin Sandwich Problem}.            
The {\it thick sandwich} prescribes knowns $h_{ij}^{(1)}$ and $h_{ij}^{(2)}$ on two hypersurfaces -- the `slices of bread' -- and one is to solve for the finite region of `filling' 
in between (Fig 1.b), in analogy with the QM set-up of transition amplitudes between states at two different times \cite{WheelerGRT}.
This turns out to be very ill-defined mathematically.
The thin sandwich is then Wheeler's \cite{WheelerGRT} `thin limit' of this, with spatial metric $h_{ij}$ and its label-time velocity $\dot{h}_{ij}$ prescribed as data on a spatial 
hypersurface $\Sigma$ (Fig 1.c).
Here one is to solve for $\beta^i$ the {\it thin sandwich equation} -- ${\cal M}_i$ in Lagrangian variables, 
including taking an emergent position \cite{BSW} on the form of the lapse, $\alpha = \sqrt{T_{\sA\sD\sM}^{\sG\sR}/4\{R - 2\Lambda\}}$:  
\beq
D_{j}\left\{ {\sqrt{\frac{2\Lambda - R}{\{h^{ac}h^{bd} - h^{ab}h^{cd}\}
                                        \{\dot{h}_{ab} - 2 D_{(a} \beta_{b)}\}
                                        \{\dot{h}_{cd} - 2 D_{(c} \beta_{d)}\}  }}} 
                                        \{h^{jk} \delta^{l}_{i} - \delta^{j}_{i}h^{kl}   \}
                                        \{\dot{h}_{kl} - 2 D_{(k} \beta_{l)}\}\right\} = 0 \mbox{ } .   
\label{Thin-San}
\eeq
From this, one constructs an infinitesimal piece of spacetime to the future of $\Sigma$ via forming the extrinsic curvature combination (\ref{K-def}).  
The Thin Sandwich Problem remains a problem because its partial differential equation mathematics is hard \cite{TSC1, TSC2}.  

\mbox{ } 

\noindent {\bf Functional Evolution Problem}.     
This concerns whether no more constraints than $\widehat{\cal H}$ and $\widehat{\cal M}_i$ are required at the quantum level. 
More might be required, since given quantum-level constraint equations
\be
\widehat{\cal C}_{\sfC} \Psi = 0 \mbox{ } \not{\Rightarrow \mbox{}} \mbox{ } \mbox{\bf [}\widehat{\cal C}_{\sfC}\mbox{\bf ,} \, \widehat{\cal C}_{\sfC^{\prime}}\mbox{\bf ]}\Psi = 0 
\label{FEP}
\ee 
automatically as well.  
Instead, more constraint terms might be unveiled, or the right-hand-side of the second equation in (\ref{FEP}) might be an anomaly term rather than zero.  
For GR in general, this remains an unsolved problem.  

\mbox{ } 

\noindent {\bf Problem of Observables}.           
This concerns finding enough quantities $\widehat{\cal O}_{\sfO}$ to describe the physics, these {\it observables} being defined as commutants with all of a theory's first-class constraints 
\beq
\mbox{\bf [}\widehat{\cal C}_{\sfF}\mbox{\bf ,} \, \widehat{\cal O}_{\sfO}\mbox{\bf ]} = 0
\eeq 
-- Dirac observables -- or maybe just with the linear ones -- \K observables.
For GR in general, this problem is, once again, an unresolved one \cite{PoB}.  

\mbox{ } 

\noindent {\bf Foliation Dependence Problem}.     
At the classical level, this concerns whether evolving via the dashed or the dotted surface in Fig 1.e) gives the same answer \cite{T73}.
GR succeeds in this way as per Sec \ref{Fol-SRP}, and this is held to be part of GR's desirable coordinate independence, so one would like for observable inner product combinations of 
wavefunctions and operators to maintain foliation independence.
Unfortunately we do not know for now how or whether this can be attained in general at the quantum level.  

\mbox{ } 

\noindent {\bf Spacetime Reconstruction Problem}. 
Reconstruction here refers to recovering spacetime from assumptions of just space and/or a discrete ontology: Fig 1.f).
It is further motivated at the quantum level, as per Fig 1.g) and h).  

\mbox{ } 

\noindent {\bf Global Problem of Time}.           
In \K and Isham's reviews, this consists of difficulties with choosing an 'everywhere-valid' timefunction (see \cite{Global} for an update); 
this could refer to being defined over all of space or over all of the notion of time itself.  

\mbox{ } 

\noindent {\bf Multiple Choice Problem}.          
This is only relevant once the quantum level is under consideration, and, as Fig 1.i) illustrates, 
consists of canonical equivalence of classical formulations of a theory not implying unitary equivalence of the quantizations of each \cite{Gotay}.  
By this, different choices of timefunction can lead to inequivalent quantum theories.
%
{            \begin{figure}[ht]
\centering
\includegraphics[width=1.0\textwidth]{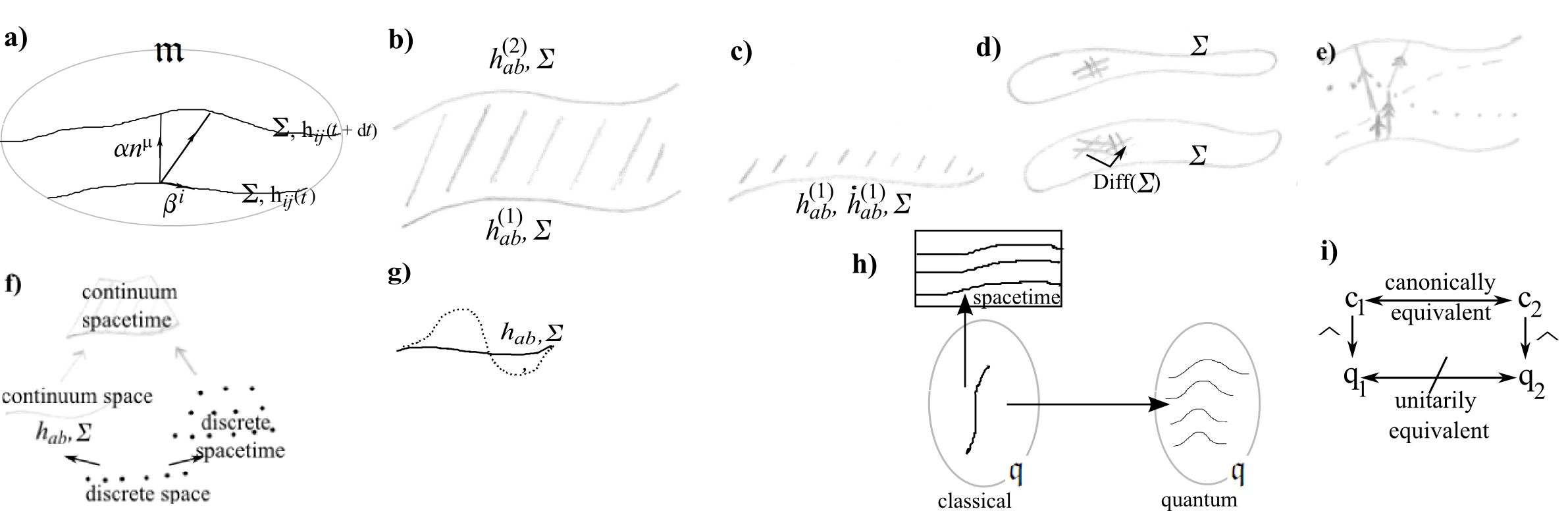}
\caption[Text der im Bilderverzeichnis auftaucht]{        \footnotesize{ a) ADM split of the spacetime metric.
b) Thick Sandwich and its Thin Sandwich limit c).  
The data are as given and the problems to solve are for the spacetime in each shaded region. 
d) is the Thin Sandwich's reworking as the geometrodynamical case of Best Matching: with respect to the spatial diffeomorphisms Diff($\Sigma$).  
e) depicts the geometry of the text's statement of the Foliation Dependence Problem.
f) to h) depict Spacetime Reconstruction issues.  
f) outlines what different levels of reconstruction assume as starting points.
g) depicts the dynamical object -- the spatial 3-geometry (solid) -- and the subsequent quantum fluctuations of this, (dotted) which do not all fit into the one spacetime.
h) Moreover, precisely-known position $\underline{q}$ and momentum $\underline{p}$ for a particle are a classical concept corresponding to a worldline.
This view of the world is entirely accepted to break down in quantum physics due to Heisenberg's Uncertainly Principle; 
in QM, worldlines are replaced by the more diffuse notion of wavepackets. 
Wheeler then pointed out \cite{Battelle, W79} that in GR, the uncertainty principle now applies to the quantum operator counterparts of $h_{ij}$ and $\pi^{ij}$. 
But by formula (\ref{Gdyn-momenta}) this means that $h_{ij}$ and $K_{ij}$ are not precisely known.   
Finally, i) supports the statement of the Multiple Choice Problem. 
Here `c' stands for classical formulation, `q' for quantum formulation and $\widehat{\mbox{ }}$ denotes quantization map.} }
\label{Facet-Intro} \end{figure}          }


\noindent Additionally, Isham and Kucha\v{r} classified strategies for the Problem of Time into the following.  

\mbox{ } 

\noindent 1) {\bf Tempus ante Quantum}.
Time exists prior to quantization, in one of the following forms.

\noindent i) Perhaps time is internal to one's gravitational theory.

\noindent ii) Perhaps time is provided by appending certain kinds of matter to the gravitational theory. 

\noindent iii) Unimodular time is the momentum conjugate to $\Lambda$, upon elevating this to a dynamical variable.

\mbox{ } 

\noindent 2) {\bf Tempus post Quantum} this involves time emerging at the quantum level in 

\noindent i) the {\it Klein--Gordon-like formulation}.
[This is named for the unfortunately only superficial \cite{Kuchar81} similarity between GR's configuration space Riem($\Sigma$) and Minkowski spacetime: 
both have indefinite metrics upon them: the inverse DeWitt supermetric for GR and the obvious Minkowski metric.]

\noindent ii) The {\it semiclassical approach}, in which some heavy slow degrees of freedom provide an approximate emergent time with respect to which the other light fast degrees of freedom evolve.
See Sec 7 for an outline and \cite{Kieferbook, FileR, ACos2} for further details.  

\noindent iii) {\it Third quantization} is the alternative suggestion then is that the solutions $\Psi[\bh]$ of the WDE might be turned into operators, so that one now has an equation
$\widehat{\cal H}\widehat{\Psi}\psi = 0$.   
[Compare the standard notion of second quantization in QFT.]

\mbox{ } 

\noindent 3) {\bf Tempus Nihil Est}, i.e. making do with no time.
Examples of this include 

\noindent i) the {\it \NSI} for answering questions about being rather than becoming. 

\noindent ii) The {\it \CPI} for answering questions of conditioned being, which can then be about being at a time or about correlations.

\noindent iii) {\it Histories Theory} \cite{GMH, Hartle, IL}, in which histories themselves are regarded as primary entities.  

\noindent iv) Approaches involving \cite{RovelliBook} partial observables 
-- which do not require commutation with any constraints, and contain unphysical information but are such that one can consider correlations between pairs of them that are physical -- 
and evolving constants of the motion.

\noindent A number of extra programs have been added since, and I gave a more extensive classification than the above in \cite{APoT2}.

\mbox{ }

\noindent I next note that \cite{FileR, APoT2} all bar one of PoT the facets have classical precursors, hence each of Secs \ref{TR}-\ref{Fol-SRP} starts to consider each facet 
at the classical level.
As the present Article progresses, it also upgrades most of the names and concepts of the facets; the outcome of this is then summarized in Sec \ref{Up-Fa}.  
The position I take is that    A) GR is a gestalt theory --- both a relativistic theory of gravitation {\sl and} an attempt to free Physics of background structure \cite{FileR}. 
                               B) Background independence is philosophically and physically desirable \cite{BI}.
Background independent theories include not only geometrodynamics but also Loop Quantum Gravity, Supergravity and M-Theory, but not perturbative String Theory itself.

\mbox{ }
							   					   
\noindent Then Barbour's work \cite{B94I} is background independent at the classical level, it leads to two of the PoT facets, 
and my extension of this work unearths classical counterparts of all PoT facets bar the purely quantum mechanical Multiple Choice Problem.
In particular, Sec \ref{TR} explains the Temporal Relationalism underpinning of the Frozen Formalism Problem, alongside the classical Machian resolution of this issue.
Also, Sec \ref{CR} explains the Configurational Relationalism generalization of the Thin Sandwich Problem.  
This approach also sheds further light on the meaning of, and strategization for, both the classical- and quantum-level PoT.
Sec 4 covers Spacetime Relationalism.
Sec 5 explains what the Functional Evolution Problem and Problem of Observables become, whereas Sec 6 expands on the already mature Foliation Dependence Problem 
and new material on the Spacetime Reconstruction Problem.
Further quantum aspects of the facets are outlined in Sec 7.  

\mbox{ } 

\noindent Moreover, \K cautioned that PoT facets resist being resolved piecemeal. 
Note that the various facets arise from a common cause \cite{FileR}: the conceptual mismatch between GR and ordinary Quantum Theory.
\K \cite{Kuchar93} compared attempting to resolve them to going through a series of gates only to find oneself outside of some of the gates one thought one had already left behind.
For instance, functional evolution can be foliation dependent, and one cannot start to find \K or Dirac observables until one has a consistently full set of constraints \cite{PoB}.
The present Article points out a number of other such interferences, both among local facets (Secs \ref{CR}-\ref{Combo}) and as regards how a large majority of facets and strategies 
having global issues (Sec \ref{Conclusion}).

\mbox{ } 

\noindent I additionally provide a local resolution to the PoT that is Machian in character (in the senses explained in Secs 2 and 3).  
The strategy is laid out as I reconceptualize and overcome each of the first seven facets that form `a local' resolution of the PoT  
(i.e. not facing the Multiple Choice or Global Problems).
I do this via a 3-level approach consisting of a Machian classical resolution \cite{ACos2}, a Machian semiclassical resolution \cite{SemiclIII, ACos2} 
and a combined Machian semiclassical, histories and records scheme \cite{AHall, FileR}.
The model arenas in which I do this are the Jacobi formulation of Mechanics \cite{Lanczos}, the relational triangle \cite{BB82, FileR} and minisuperspace GR \cite{AMSS1, AMSS2}, 
with an outline of the extension of this to so far the classical part of slightly inhomogeneous cosmology \cite{SIC}.

\section{Temporal Relationalism underlies the Frozen Formalism Problem}\label{TR} 

The new conceptual starting-point is {\bf Temporal Relationalism}.
This consists of adopting Leibniz's `there is no time for the universe as a whole' principle \cite{B94I, FileR} as a desirable tenet of background independence and of closed universes.
This is mathematically implemented by postulating actions that 

\noindent i)  do not contain any extraneous time (such as Newton's) or time-like variables (such as GR's lapse). 

\noindent ii) They are {\sl geometrical Jacobi--Synge type actions} that happen to be parametrization-irrelevant.  

\noindent [This is a conceptual evolution of considering first reparametrization-invariant actions and then parametrization-irrelevant ones that do not even involve a parameter. 
Moreover, the logical conclusion of this process is to neither name nor conceive in terms of what is not present in these actions. 
These actions always had a geometrical character as well, and this aspect of them is retained and thus ends up being the most apt conceptualization and name for them.] 

\mbox{ } 

\noindent Examples of such actions are Jacobi's principle \cite{Lanczos} for Mechanics or Misner's parageodesic principle \cite{Magic} for minisuperspace GR.
Both are of the form\footnote{The background independent formulation of Mechanics already possesses \cite{FileR} 6 of the 8 facets of the canonical PoT.  
This renders it a useful model arena for quite a few PoT investigations.  
This study is to be complemented with models that nontrivially involve diffeomorphisms and GR spacetime-like notions. 
This is since the hitherto missing 2 facets are of that nature.  
Also, when the Configurational Relationalism involves diffeomorphisms, this renders most of the other facets more technically complicated as well.}
\beq
S = \sqrt{2}\int\d s\sqrt{W(\mbox{\boldmath$Q$})} =: \int \d J \mbox{ } . 
\eeq 
Here $W$ is the potential factor. 
For mechanics, this takes the form $W = E - V$                                            for total energy   $E$ and potential energy        $V$.
On the other hand, in (for now minisuperspace) GR, this takes the form $W = R - 2\Lambda$.   
Also $\d s := ||\d \mbox{\boldmath$Q$}||_{\sbM}$ is the kinetic arc element (configuration space geometry with metric $\bM$). 
$\d J$ is the conformally-related physical line element (the conformal factor being $\sqrt{2W}$).
Thus this action principle is a {\it geodesic principle} in $\d J$ or a {\it parageodesic principle} \cite{Magic} in $\d s$ (i.e. geodesic up to a conformal factor).  
We finally note \cite{FileR} equivalence to the more common Euler--Lagrange or ADM equations by moves such as passage to the Routhian \cite{Lanczos} or Lagrange multiplier elimination.   

\mbox{ } 

\noindent Next, Dirac \cite{Dirac} noted that primary constraints are implied by reparametrization-invariant actions. 
[Hence this is also holds for our conceptually-enhanced equivalent of these.]
This accounts for how action for minisuperspace GR manages to encode the Hamiltonian constraint ${\cal H}$.  
In the ADM approach, this arises instead by variation of the lapse, which is itself absent from Misner's action.
Thus the constraint whose quadratic dependence on the momenta is well-known to cause the Frozen Formalism Problem arises directly from the demand of Temporal Relationalism. 
Its precise form is dictated by the way the action is built to be temporally relational.
Thus indeed Temporal Relationalism is a deeper and already classically-present replacement for the Frozen Formalism Problem.  
For Jacobi's formulation of mechanics, the quadratic energy constraint ${\cal E} := ||\mP||_{\mbox{\scriptsize \boldmath $N$}}\mbox{}^2 + V = E$ 
(for $\mbox{\boldmath $P$}$ conjugate to $\mbox{\boldmath $Q$}$) plays an analogous role to GR's ${\cal H}$.

\mbox{ } 

\noindent Moreover, the above primary-level timelessness can be resolved at a secondary, emergent level by {\it Mach's Time Principle}: `time is to be abstracted from change'. 
Three distinct specifications of this involve  `any change' (Rovelli \cite{Rfqxi}), 
`all change' (Barbour \cite{Bfqxi}) and my {\it sufficient totality of locally significant change} ({\it STLRC}) \cite{ARel2}.  
This emergent time represents a local generalization of the astronomers' ephemeris time; this is particularly manifest in the case of mechanics.
Generalized local ephemeris time is to be abstracted from STLRC.
To fulfil the true content of the STLRC approach, all change is given opportunity to contribute to the timestandard. 
However only changes that do so in practise to within the desired accuracy are actually kept.

\mbox{ } 

\noindent For the actions in question, emergent Jacobi time resolves Mach's Time Principle, at first sight in the `all change' manner, but, in practice in the `STLRC' manner.
It is, furthermore, a simplifier of the change-momentum relations and Jacobi--Mach equation of motion 
(temporally relational equivalents of velocity-momentum relations and Euler--Lagrange equations \cite{AM13}).  
A general formula for this is (using `J' to denote `Jacobi')
\beq
\lt^{\se\sm(\sJ)}             =            \int \d s\left/\sqrt{2 W(\mbox{\boldmath$Q$})}\right. \mbox{ } .
\eeq
Here the oversized notation $\mbox{\Large $t$}^{\se\sm(\sJ)} := t^{\se\sm(\sJ)} - t^{\se\sm(\sJ)}(0)$ is used to incorporate selection of `calendar year zero', 
$t^{\se\sm(\sJ)}(0)$. 
%
%
\noindent The above amounts to a relational recovery of Newtonian, proper and cosmic time in suitable contexts.

\mbox{ } 

\noindent In the presence of an h--l split (heavy slow degrees of freedom and slow fast ones), as is the case for Cosmology, 
this scheme is only fully Machian once one passes from the zeroth-order emergent times whose form is F[h, dh] to at least-first order emergent times of from F[h, l, dh, dl].  
I.e. giving the l degrees of freedom the opportunity to contribute.

\noindent  The above classical resolution does not produce a timefunction that carries over at the quantum level. 
However, as we shall see in Sec \ref{Semi}, there is a quantum-level emergence that parallels the above classical emergence.

\section{Configurational Relationalism generalizes the Thin Sandwich Problem}\label{CR}

\noindent The new starting point involves considering configuration space $\FrQ$ and then a group $\FrG$ of continuous transformations that are taken to be physically irrelevant.
This encompasses both of the following. 

\noindent 1) Spatial  Relationalism [translations and rotations relative to absolute space in Mechanics, or Diff($\Sigma$) in GR].

\noindent 2) Internal Relationalism [a reformulation of the more familiar type of gauge theories from Particle Physics].

\noindent This can be implemented indirectly in a very wide range of circumstances, by the following `$\FrG$-act $\FrG$-all' method. 
Given an object $O$ that corresponds to the theory with configuration space $\FrQ$, one first applies a group-action of $\FrG$ to this --- denoted $\stackrel{\rightarrow}{\FrG}_gO$. 
Then one applies some operation $S_g$ that makes use of all of the $g \in \FrG$ so as to cancel out the appearance of $g$ in the group action, 
e.g. summing, integrating, averaging or extremizing over $\FrG$.  

\noindent One example of this, for $O$ a classical action built upon $\FrQ$, is to apply the basic infinitesimal group action to obtain
\beq
S_{\mbox{\scriptsize relational}} = \sqrt{2} \int \int_{\sN\so\sS} \d (\mN\mo\mS) \d_g s \sqrt{W(\mbox{\boldmath$Q$})}  
\mbox{ } , \mbox{ } \mbox{ } 
\d_gs := ||\d \mbox{\boldmath$Q$} - \stackrel{\rightarrow}{\FrG}_{\d g} \mbox{\boldmath$Q$}||_{\sbM}  
\eeq
and then to extremize over $\FrG$ as per the variational principle now also including variation with respect to $g$.\footnote{`NoS' denotes each configurational entity's notion of space: 
3-space for field theories, whilst, for finite theories, we take $\int_{\sN\so\sS} \d (\mN\mo\mS) := 1$.}
This particular example is Barbour's {\sl Best Matching} \cite{ARelB11GrybTh}.
This name emphasizes the bringing into maximum congruence of the adjacent configurations. 
The GR subcase of this example is indeed the Thin Sandwich Problem facet \cite{FileR} of  Fig 1.c)  [NoS = $\Sigma$, $\FrG = \mbox{Diff}(\Sigma)$], 
with corresponding action \cite{RWR, FileR, AM13}
\beq
S^{\sG\sR}_{\sr\se\sll\sa\st\si\so\sn\sa\sll} = \int \int_{\Sigma} \d^3x \sqrt{h} \d_{\underline{F}}s \sqrt{R - 2\Lambda} 
\mbox{ } \mbox{ } \mbox{ , }  \mbox{ }  
\d_{\underline{F}} s := ||\d_{\underline{F}} \mbox{\boldmath$h$}||_{\mbox{\boldmath\scriptsize $M$}}
\mbox{ } \mbox{ } \mbox{ and }  \mbox{ } 
\d_{\underline{F}} h_{ij} :=  \d h_{ij} - \pounds_{\underline{F}}h_{ij} 
\mbox{ } .  
\eeq
Here $\underline{F}$ is a Diff($\Sigma$) auxiliary vector, such that $\dot{\underline{F}}$ is the more conventional formulation's $\underline{\beta}$.

Thus Configurational Relationalism is  a twofold generalization.
I.e. firstly to Best Matching (from GR to a wide range of theories) and secondly to encompass tackling the physically redundant group at any level of structure, rather than 
specifically at the Lagrangian level.

As another example of Best Matching, see Fig 2.a) for a (3-particle, scaled) relational particle mechanics (RPM) \cite{FileR} 
case of this [NoS trivial, $\FrG$ the rotations in the relational coordinates formulation], for which the action (for $N$ particles) is \cite{FileR} 
\beq
S^{\sR\sP\sM}_{\tr\te\tl\ta\ttt\ti\to\tn\ta\tl} = \sqrt{2}\int\d_{\underline{B}}s\sqrt{E - V(\brho)} 
\mbox{ } , \mbox{ } \mbox{ } \mbox{ and }  \mbox{ } 
\d_{\underline{B}}s := || \d\brho - \d \underline{B} \cr \d\brho ||  \mbox{ } .
\eeq
Here $\brho$ are relative Jacobi coordinates as exemplified in Fig 2, 
and $\underline{B}$ is the rotational auxiliary vector (which has only one component in the 2-$d$ case we focus on below).  
The corresponding constraint is the zero total angular momentum constraint ${\cal L}_i := \sum_{I = 1}^{N - 1} \underline{\rho}^I \cr \underline{\pi}_I = 0$ 
for $\underline{\pi}_I$ the momenta conjugate to $\underline{\rho}^I$.  

\mbox{ } 

\noindent The extremization produces an equation that, in the $\mbox{\boldmath $Q$}, \d \mbox{\boldmath $Q$}$ variables formulation, is to be solved for $g$ itself 
and then substituted back into the action. 
This produces a final $\FrG$-independent expression that directly implements Configurational Relationalism.  
Moreover, the initial indirectly formulated expression
\beq
\lt^{\se\sm(\sJ\sB\sB)} = \mbox{\Large E}_{g \in \sFG}                                                      
\left(                                                              
\int||\d_{g} \mbox{\boldmath$Q$}||_{\mbox{\scriptsize \boldmath $M$}}\left/\sqrt{2 W(\mbox{\boldmath$Q$})} \right. 
\right)
\label{t-JBB} 
\eeq
itself succeeds in implementing Configurational Relationalism.  
Here $\mbox{\Large E}_{g \in \sFG}$ denotes `extremum of $g \in \FrG$ of $S_{\sr\se\sll\sa\st\si\so\sn\sa\sll}$ built upon $\FrQ, \FrG$.   
This is because, whilst it initially extends $\FrQ$ to the bundle $P(\FrQ,\FrG)$ by including $g$-auxiliaries, 
$g$-variation then gives a gauge constraint and these use up 2 degrees of freedom per $g$, so one indeed ends up on the quotient space $\FrQ/\FrG$ as required.  
(JBB stands for `Jacobi--Barbour--Bertotti' \cite{BB82}.)  

\noindent Note that the expression given involves formulating the actually-present auxiliary variables as $\d g$.  
This is necessary \cite{FEPI, FileR} for these not to spoil the parametrization-irrelevance that implements Temporal Relationalism.

\noindent [Solving for a cyclic differential of a frame variable $\d F^i$ in place of a multiplier coordinate shift $\beta^i$, 
as required \cite{AM13} for compatibility with Temporal Relationalism, in no way alters the mathematics of the thin sandwich equation (\ref{Thin-San}).]

\mbox{ } 

\noindent The scaled relational mechanics and GR cases of (\ref{t-JBB}) are, respectively, 
\beq
\lft^{\se\sm(\sJ\sB\sB)} = \mbox{\Large E}_{\underline{B} \mbox{ } \in \mbox{ } \mbox{\scriptsize Rot}(d)}                                                                                                                   
\left(                                                              
\int \d_{\underline{B}} s\left/\sqrt{E - V(\brho)} \right. 
\right)                                                                         \mbox{ } ,
\eeq
\beq
\lt^{\se\sm(\sJ\sB\sB)} = \mbox{\Large E}_{\underline{F} \in \mbox{\scriptsize Diff}(\Sigma)}
\left(
\int    \d_{\underline{F}} s\left/\sqrt{R - 2\Lambda}\right.
\right)                                                                         \mbox{ } .
\eeq
%
{            \begin{figure}[ht]
\centering
\includegraphics[width=0.8\textwidth]{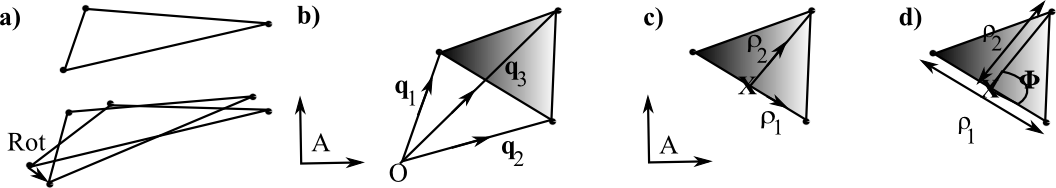}
\caption[Text der im Bilderverzeichnis auftaucht]{        \footnotesize{a) is Barbour's well-known Best Matching of the relational triangle \cite{EOT} (RPM analogue of the thin sandwich). 
The rest of this figure is a progression of coordinate systems for the relational triangle. 
b) are particle position coordinates relative to an absolute origin O and absolute axes A.  
c) are relative Jacobi interparticle cluster separations; X denotes the centre of mass of particles 1 and 2; note that these coordinates are still relative to absolute axes A.
Then the configuration space radius $\rho := \sqrt{\rho^2_1 + \rho^2_2}$.  
d) are scaled relational coordinates (ie no longer with respect to any absolute axes either).  
Pure-shape coordinates are then the relative angle $\Phi$ and some function of the ratio $\rho_2/\rho_1$; in particular, $\Theta := 2\,\mbox{arctan}(\rho_2/\rho_1)$.} }
\label{RPM-coordi} \end{figure}          }
%
\noindent \mbox{ } \mbox{ } Moreover, the former's Best Matching is explicitly solvable in 1- and 2-$d$ \cite{FileR, QuadI}.  
By use of Kendall's Shape Theory \cite{Kendall} and the coning construction \cite{FileR}, the simplest configuration space geometries for 1- and 2-$d$ RPM's are $\mathbb{S}^{n - 1}$ 
and $\mathbb{CP}^{n - 1}$ (for pure shapes, i.e. models free of scale), and $\mC(\mathbb{S}^{n - 1}) = \mathbb{R}^{n}$ and $\mC(\mathbb{CP}^{n - 1})$ (for models also including scale). 
The first three of these are very well known as geometries and as regards subsequent classical and quantum mechanics thereupon and supporting linear methods of Mathematical Physics.  
These render many QM and PoT strategy calculations tractable and available for comparison with each other, which is a rarity in the latter field.  
Triangleland is further aided in this way by $\mathbb{CP}^{1} = \mathbb{S}^{2}$ and $C(\mathbb{CP}^1) = \mathbb{R}^3$ albeit the latter is not flat; it is, however, conformally flat.
The simpleness of the ensuing mathematics, even well into the usually complex rearrangements necessary for the investigation of PoT strategies, is a major asset in this RPM model 
arena. 
This is because it secures many computational successes beyond the usual points at which these break down for full GR/many other model arenas.  
Pure-shape RPM configuration spaces are analogous to conformal superspace (CS) for GR, and scaled ones to Wheeler's superspace in one sense and to CS + Volume \cite{ABFKO} in another.
\cite{AF+tri, QuadIII} demonstrated solvability. 
This is by a mixture of basic maths and interdisciplinarity with statistical theory of shape, 
Molecular Physics and a few other areas (Particle Physics, instantons) for the quadrilateralland and higher-$N$ $N$-a-gon cases.  
\noindent RPM isometry groups have 1) Atomic/Molecular Physics analogies: $SO(3) = SU(2)/\mathbb{Z}_2$ for triangleland.  
2) Particle Physics analogies: the $SU(3)$/$\mathbb{Z}_3$ for quadrilateralland \cite{QuadI} 
is identical to the colour group and shares the same Lie algebra with approximate flavour physics as well.  

\mbox{ }

\noindent Scaled triangleland has non-obvious Cartesian coordinates -- {\it Dragt coordinates} -- that are useful below,
\beq
Dra_1 = 2 \,  \underline{\rho}_1 \cdot \underline{\rho}_2     \mbox{ } , \mbox{ } \mbox{ }
Dra_2 = 2\{ \underline{\rho}_1  \cr  \underline{\rho}_2\}_3 \mbox{ } , \mbox{ } \mbox{ }
Dra_3 = \rho_2\mbox{}^2 - \rho_1\mbox{}^2 \mbox{ } .  
\label{Dragt}
\eeq
These are all cleanly interpretable as the product of a scale factor $I$ and a lucid shape quantity.
I.e. a departure from isoscelesness, 
four times the mass-weighted area of the triangle (the 3 denotes component in the fictitious third dimension out of the plane from the triangle), 
and the ellipticity (difference of partial moments of inertia) respectively. 
(\ref{Dragt}) are also closely related to the Hopf map $\mathbb{S}^3 \rightarrow \mathbb{S}^2$.

\section{Constraint closure and expression in terms of beables}\label{CC-Beables}

As well as $\mbox{\boldmath $Q$}$'s and $\mbox{\boldmath $P$}$'s one requires the Poisson bracket, $\mbox{\bf \{} \mbox{ } \mbox{\bf ,} \mbox{ } \mbox{\bf \}}$. 
Then the constraints have brackets among themselves.

The Problem of Time account's Functional Evolution Problem at QM level is for field theories (functional as in `functional derivative').    
One has, rather, a Partial Evolution Problem for finite theories and then the portmanteau of these two sorts of derivative the general case covering both of these: 
the `Partional Evolution Problem' \cite{FileR}.   
However, `Constraint Closure Problem' is still stronger as a concept and name, since it clearly applies at the classical level too. 
Thus I use and recommend that name for the third Facet.

The Functional Evolution Problem is viewed as part of the a posteriori compatibility for relational models, and is fortunately absent in this Article's RPM's. 
I generalize this to simply the Constraint Closure Problem so as to include the classical case which the Dirac algebroid's closure indeed resolves for classical GR.
Foliation-independence is also classically guaranteed by the Dirac algebroid (Fig 3).
As regards the Spacetime Reconstruction Problem, space/configurations/dynamics are primary, and spacetime may not exist as a meaningful concept at the level of Quantum Gravity.

As regards some examples, the only nonzero constraint Poisson brackets for RPM is 
\beq
\mbox{\bf \{}    {\cal L}_i    \mbox{\bf ,} \, {\cal L}_j    \mbox{\bf \}} = \epsilon_{ij}\mbox{}^k{\cal L}_k \mbox{ } .  
\eeq
For the GR case, see the lower box in Fig \ref{DiracAlgebroid2} for the Dirac algebroid of the GR constraints \cite{Dirac}.

\mbox{ } 

\noindent Given the brackets, one can ask about which objects (observables/beables) have 
%
%
zero brackets with the (first-class) constraints too.
The problem is that (a sufficient set of) these are hard to come by in classical and quantum gravitation.
The distinction between observables and beables, and reason for the name change from Problem of Observables to Problem of Beables, is as follows \cite{Bell75}. 
It is the difference between entities being observed and entities simply being, so the circumstances under which observables occur are then a subset of those in which beables do.  
Moreover, from a beables perspective, defining what `observing' is is unnecessary, so conceptualizing in terms of beables is a freeing from having to define this.
Two contexts in which beables are relevant are then 1) whole-universe or closed-system modelling \cite{Bell}.
%
%
2) At the quantum level, where the connection between the notion of observation and the quantum Measurement Problem \cite{Measurement}.

Trivial Configurational Relationalism (or resolved Best Matching) readily imply possession of a full set of classical \K beables, 
i.e. quantities that Poisson-brackets-commute with the classical linear constraints.    
For the relational triangle, these are 
\beq
K = F[\mbox{\boldmath $Dra$}, \mbox{\boldmath $P$}^{Dra} \mbox{ alone}] \mbox{ } , 
\eeq
for  $\mbox{\boldmath $P$}^{Dra}$ the conjugate variables to $\mbox{\boldmath $Dra$}$.
In the case of trivial Configurational Relationalism, Halliwell provided \cite{H03, H09H11} a prescription for Dirac beables -- commuting with the quadratic constraint also -- 
which I promoted to the case of resolved Best Matching too \cite{AHall}.
\noindent The Problem of Beables consists of finding objects which brackets-commute with all the constraints (Dirac beables) or perhaps just with the linear constraints.
Dirac beables are sufficiently hard to find for full GR that \K \cite{Kuchar93} likened postulating having obtained a full set of these to having a unicorn as one's willing steed.

Consequences of Best Matching Problem Resolution are 1) automatic availability of classical \K beables. 
2) The classical Constraint Closure Problem is then resolved by there being only one constraint (per space point in field-theoretic case) 
-- the reduced formulation's quadratic constraint -- which then straightforwardly closes with itself.  
This often relies on the $\FrG$ being an ultimately compatible choice for the $\FrQ$ in question, no extra integrabilities, no extra QM constraints and no anomalies.

\noindent I distinguish between specific and merely formal resolutions of the Problem of \K Beables. 
E.g. for the triangle, one has a specific set of shape quantities, whereas for GR one can only talk formally in terms of the spatial 3-geometries.   

\mbox{ } 

\noindent Triangleland's three other classical facets are resolved by foliations and spacetime not being meaningful concepts in this arena, 
and by straightforward computation of the constraint algebra.

\section{Spacetime Relationalism}\label{SpR}

GR has more background independent features than Mechanics theories do.
This is due to GR having a spacetime notion, which has more geometrical content than Mechanics' space-time notion does.
Spacetime's own relationalism is then characterized as follows. 


\noindent i) The are to be no extraneous spacetime structures, in particular no indefinite background spacetime metrics. 
Fixed background spacetime metrics are also more well-known than fixed background space metrics. 

\noindent ii) Now as well as considering a spacetime manifold $\FrM$, consider also a $\FrG_{\sS}$ of transformations acting upon $\FrM$ that are taken to be physically redundant.


\noindent For GR, note that $\FrG_{\sS}$ = Diff($\FrM$).
\noindent Also note that i) and ii) can be extended to include no extraneous internal structures now viewed as fields on spacetime, 
with $\FrM$ being extended to a product with internal spaces and $\FrG_{\sS}$ acting upon this product space.
The internal part of ii) is closer to the most commonplace presentation of gauge theory than the spatial part of Configurational Relationalism is, 
since that is also a spacetime presentation.
On the other hand, the most commonplace presentation of gauge theory is more closely tied to Dirac observables/beables, out of these two things both being configuration-based notions.

Diff($\FrM$) indeed straightforwardly forms a Lie algebra, in parallel to how Diff($\Sigma$) does:   
\be
\mbox{\bf |[} 
(    \mD_{\mu}    |    X^{\mu})    
\mbox{\bf ,} \, 
(    \mD_{\nu}|Y^{\nu}    ) 
\mbox{\bf ]|} = 
(    \mD_{\gamma}    | \, [X, Y]^{\gamma}    ) \mbox{ } \label{Lie-2} . 
\ee
Here the D's are generators, $\mbox{\bf |[} \mbox{ }  \mbox{\bf ,} \, \mbox{ }  \mbox{\bf ]|}$ is a generic Lie bracket, 
                                   and $[ \mbox{ } , \mbox{ } ]$ is the differential-geometric commutator.
Diff($\FrM$) also shares further specific features with Diff($\Sigma$), such as its right hand side being of Lie derivative form.

However, whereas Diff($\Sigma$) generators are conventionally associated with dynamical constraints, Diff($\FrM$)'s are not. 
Additionally, Diff($\Sigma$)'s but not Diff($\FrM$)'s classical realization of the Lie bracket is conventionally taken to be a Poisson bracket.
This furthermore implies that there is conventionally no complete spacetime analogue of the previous Chapter's notion of beables/observables.
These differences are rooted in time being ascribed some further distinction in dynamical and then canonical formulations than in spacetime formulations.   
(\ref{Lie-2}) is to be additionally contrasted with the Dirac algebroid in the second box of Fig \ref{DiracAlgebroid2}.  
Clearly there are two very different algebraic structures that can be associated with GR spacetime: 
the first with unsplit spacetime and the second with split space-time including keeping track of how it is split.

Diff($\FrM$) is closely related to {\it spacetime observables} in GR.
Such objects would be manifestly Diff($\FrM$)-invariant, i.e. commutants $\mS_{\sfQ}$:  
\be
\mbox{\bf |[} (    \mD_{\mu}    |    X^{\mu})  \mbox{\bf ,} \, (    \mS_{\sfQ}|Y^{\sfQ}    ) \mbox{\bf ]|} \mbox{ }  `=' \mbox{ } 0 \mbox{ } \label{Sp-Obs} .
\ee
 
\noindent Also note that Configurational Relationalism and \K beables involve Dirac's notion of gauge for data at a given time, 
whereas Bergmann introduced another notion of gauge for whole paths/dynamical trajectories \cite{Bergmann61}. 
Bergmann's notion of gauge leads to a distinct notion of observables, that are additionally local and independent of the Hamiltonian formalism.
A path in geometrodynamical phase space corresponds to sweeping out a slicing of a spacetime, encoding both the sequence of 3-metrics and the extrinsic curvature of each of these.

Moreover, Bergmann and Komar \cite{BK72} noted that these have invariances much larger than Diff($\FrM$), and their position is that the largest group of invariances is to be taken. 
Note that their $S^{\sB\sK}_{\sfM}$ arising thus distinct from $S_{\sfF}$ by their connection to a larger group 
-- the diffeomorphism-induced gauge group -- in place of Diff($\FrM$) itself. 
See \cite{PSS1, PSS2} for further consideration of observables from this perspective. 
Also, use of gauge generators rather than individual constraints in connection with observables is motivated and carried out in e.g. \cite{PPP, PSS1}.

\section{Foliation Dependence and Spacetime Reconstruction Problems}\label{Fol-SRP}

The first of these maintains the status quo, and the second has more recent updates.

\mbox{ }

\noindent
The split-diffeomorphism alias hypersurface-deformation algebroid (lower box in Fig \ref{DiracAlgebroid2}) is very different from the algebra of spacetime diffeomorphisms (\ref{Lie-2}).
The upper box in Fig \ref{DiracAlgebroid2} shows how the pictorial form of this algebroid implies the Refoliation Invariance resolution of the classical Foliation Dependence Problem
\cite{T73, HKT76}.
Hojman-Kucha\v{r}-Teitelboim \cite{HKT76}'s additionally obtained ${\cal H}$ from less assumptions via the nature of the deformation algebroid manifestation of the Dirac algebroid.  
%
{            \begin{figure}[ht]
\centering
\includegraphics[width=0.98\textwidth]{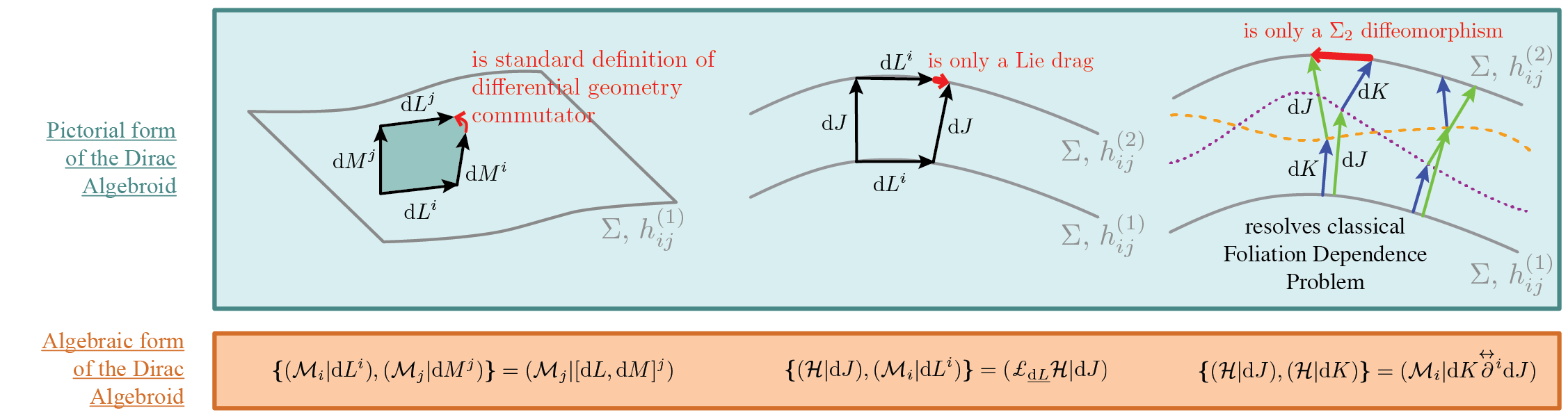}
\caption[Text der im Bilderverzeichnis auftaucht]{        \footnotesize{The Foliation Dependence Problem (Fig 1.e) is avoided for full GR at the classical level by the third figure.
$\d J$, $\d K$ are smearing functions associated with ${\cal H}$ and $\d L^i$, $\d M^i$ are smearing functions associated with ${\cal M}_i$.
Note that the smearing functions are here formulated in this differential format so as to be compatible with Temporal Relationalism \cite{AM13}.} }
\label{DiracAlgebroid2} \end{figure}          }

\noindent On the other hand Barbour--Foster--O Murchadha and I \cite{RWR} obtained ${\cal H}$ from even less assumptions.  
As I then showed with Mercati \cite{AM13}, by this stage, classical spacetime is being deduced, rather than assumed, 
from the assumption of just the geometrical structure of the space continuum, so it is a classical resolution of that aspect of the Spacetime Reconstruction Problem.  
%
{            \begin{figure}[ht]
\centering
\includegraphics[width=0.9\textwidth]{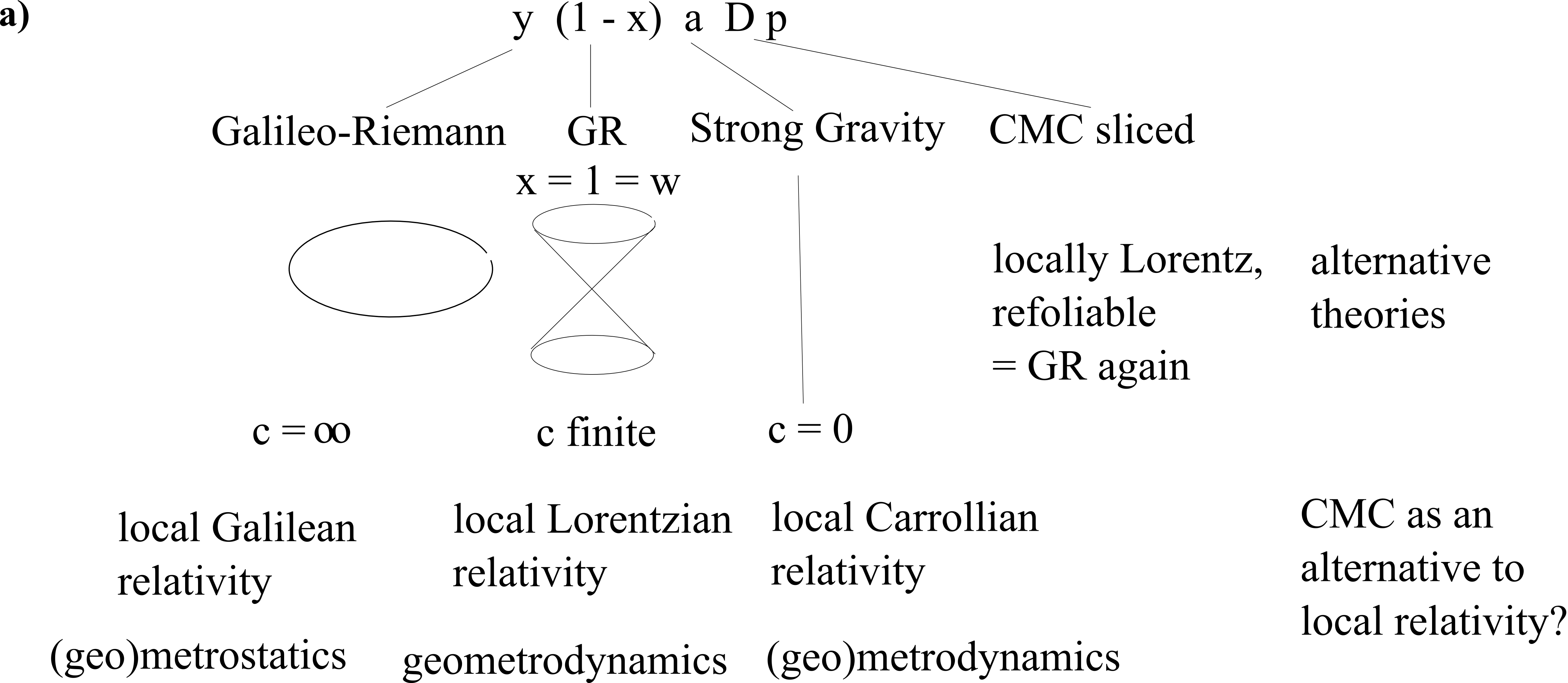}
\caption[Text der im Bilderverzeichnis auftaucht]{\footnotesize{a) If one starts with the family of theories following from the relational action $S^{w,y,a,b} = \int \int_{\Sigma} \d^3x 
\sqrt{\sqrt{h}\{a R + b\}} \, \d s_{w,y}$ with $\d s_{w,y}$ built out of $\mbox{\scriptsize \boldmath $M$}_{w,y}$ with components $M^{abcd}_{w,y} := \sqrt{h}\{h^{ac}h^{bd} - 
w h^{ab}h^{cd}\}/y$, then the Poisson brackets algebroid of the ensuing constraints \cite{AM13} yields an obstruction term with the 4 factors indicated at the top of the figure. 
[$x$ is the corresponding coefficient in the inverse supermetric.]
The figure then lays out which theories each factor leads to, how they are to be interpreted as theories of geometry and what local relativities ensue in each case upon 
inclusion of minimally-coupled matter.
In the GR case (second factor) the constraints additionally form contractions of the Gauss--Codazzi embedding equations, pointing to the existence of a surrounding 4-manifold geometry 
-- spacetime.
Some of the fourth factor's cases are known as `shape dynamics' \cite{ShapeDyn}.
Metrodynamics and metrostatics refer to versions that are free from Diff($\Sigma$) invariance. }   }
\label{AM13-2} \end{figure}          }

\section{Machian Semiclassical Resolution}\label{Semi}

{\bf Kinematical quantization}. Select a set of classical objects that are to be promoted to quantum operators, and pass from a 
classical Poisson bracket algebra to some commutator algebra that is not necessarily isomorphic to it \cite{I84}. 
In the case of the relational triangle, this is the Dragt coordinates, their conjugates and three $SO(3)$ quantities that are physically a mixture of relative angular momenta 
and relative dilational momenta \cite{FileR}.  

\mbox{ }  

\noindent{\bf Dynamical Quantization} then involves promoting the energy constraint ${\cal E}$ (or GR's ${\cal H}$) 
to a functional of the kinematical quantization operators; this procedure yields a wave equation. 
One also requires here an inner product, so as to construct observable quantities from the wavefunctions solving the wave equation.  
Dynamical quantization is the part of quantization that most concerns this article.
The wave equation in question is built on the relationally-motivated \cite{Banal} conformal ordering \cite{Magic}.  
For triangleland, moreover, either the 2-$d$ness or the flatness of the configuration space suffices to ensure that the conformal ordering is equal to the Laplacian ordering. 
Thus the wave equation is the model arena's analogue of the Wheeler--DeWitt equation, and takes the form 
\beq
-\hbar^2\{\pa^2_{\rho} + 2\rho^{-1}\pa_{\rho} + \rho^{-2}\{\triangle_{\mathbb{CP}^1} - 3/2\}\}\Psi = 2\{E_{\sU\sn\si} - V(\rho, \Theta, \Phi)\}\Psi \mbox{ } .  
\label{Gilthoniel}
\eeq
Here $E_{\sU\sn\si}$ is the energy of the model universe, taken to be fixed.

\subsection{Machian semiclassical approach's emergent time}

\noindent There is the salient problem that this Machian classical emergent time does not unfreeze the physics at the quantum level.
The way out is the Semiclassical Approach; as we shall see, this can be interpreted in Machian terms too.
We make an h--l split \cite{HallHaw, Kieferbook}.  
This can be seen as 1) a procedure from Molecular Physics by which one solves for the electronic structure under the approximation that the much heavier nuclei stay fixed, 
2) A technically similar approximation procedure from Semiclassical Quantum Cosmology.  
The Semiclassical Quantum Cosmology case's further feature is that the h degrees of freedom provide an approximate timestandard with respect to which the l degrees of freedom evolve.    
First, the h--l split is reflected at the quantum level by the wavefunction ansatz $\Psi(\mh, \ml) = \psi(\mh)|\chi(\mh, \ml)\rangle$ 
Additionally one needs to apply the WKB ansatz $\psi(\mh) = \mbox{exp}(iS(\mh)/\hbar)$ in order for this emergent time method to work.
One next considers a h-equation $\langle\chi| \times$ TISE and the l-equation $1 - |\chi\rangle\langle\chi| \times$ TISE. 
If stripped of all its quantum-mechanical terms, this becomes a Hamilton--Jacobi equation.  
This can be solved for an emergent time which coincides with the classical expression from the last section (once evaluated under the corresponding h--l split).

The l-equation looks a priori like a fluctuation equation, but becomes a time-dependent wave equation for the l-subsystem with respect to the emergent time provided by the h-equation.
If, as is usually the case, all h-derivatives bar the one in the cross-term $i\hbar\pa_{\sh}S\pa_{\sh}|\chi\rangle$ are neglected, this is a time-dependent Schr\"{o}dinger equation, 
\beq
i\hbar  \, \frac{\pa|\chi\rangle}{\pa t^{\se\sm(\sW\sK\sB)}} = \widehat{H}_{\sll} |\chi\rangle    \approx  
\frac{\hbar^2}{2}  \frac{\triangle_{\sll}}{\mh^2}   |\chi\rangle    +    V |\chi\rangle  \mbox{ }
\label{TDSE2}
\eeq
via 
\beq
i\hbar  \frac{\pa W}{\pa \mh}               \frac{\pa \left| \chi\right \rangle}{\pa \mh} = 
i\hbar \, p_{\sh}                           \frac{\pa \left| \chi\right \rangle}{\pa \mh} =
i\hbar \frac{\pa \mh}{\pa t^{\se\sm(\sW\sK\sB)}}    \frac{\pa \left| \chi\right \rangle}{\pa \mh} = 
i\hbar                                            \frac{\pa \left| \chi\right \rangle}{\pa t^{\se\sm(\sW\sK\sB)}}    \mbox{ } .  
\eeq

\noindent (\ref{TDSE2}) is, modulo the h--l coupling term, `ordinary relational l-physics'.  
The purported simple situation has `the scene set' by the h-subsystem for the l-subsystem to have dynamics. 
This dynamics is furthermore slightly perturbed by the h-subsystem, while neglecting the back-reaction of the l-subsystem on the h-subsystem.  
One might even argue for the interaction term to be quantitatively negligible as regards the observed l-physics. 

\noindent Whilst the zeroth approximation above coincides with the classical zeroth approximation, which was already declared to be non-Machian, including further correction terms 
does render the scheme Machian as follows.
Expanding the h-equation via binomial and $\hbar$ expansion moves to isolate what will often serve as first correction terms, 
\beq
\lt^{\se\sm(\sr\se\scc)} = \lt^{\se\sm(\sr\se\scc)}_{(0)} + \frac{1}{2\sqrt{2}}\int\frac{\langle  J  \rangle}{W_{\sh}^{3/2}}\frac{\d \mh}{\mh^2} - 
\frac{i\hbar}{4}  \int   \frac{\d \mh}{h^2 W_{\sh}}  \left\{  \frac{1}{\mh} + 2\left\langle \frac{\pa}{\pa \mh} \right\rangle \right\} + O(\hbar^2) \mbox{ } .   
\label{QM-expansion}
\eeq
%
Here $\langle O \rangle$ denotes the expectation $\langle \chi | O | \chi \rangle$.  

\mbox{ }

\noindent Note 1) In (\ref{QM-expansion}) I have made the `rectifying' change of variables 
\beq
\lt^{\se\sm(\sr\se\scc)} := \int\d t^{\se\sm(\sW\sK\sB)}/\mh^2(t^{\se\sm(\sW\sK\sB)})
\eeq
simplifies the l-time-dependent Schr\"{o}dinger equation (\ref{TDSE2}). 
One can then cast the h-equation in terms of this also, so as to have the entire h--l system upon a common footing as regards choice of variables \cite{ACos2, QuadIII}.   

\noindent Note 2) one can see that the difference between this Machian semiclassical emergent time and its classical counterpart is itself Machian.  
I.e. if all change is to be given the opportunity to contribute, then {\sl quantum} change is somewhat different from classical change.  
Namely, one passes from an emergent Machian time of the form $F[\mh, \ml, \d \mh, \d \ml]$ to one of the form $F[\mh, \ml, \d \mh, |\chi(\mh, \ml)\rangle]$. 
The latter takes into account that the l-subsystem has passed from a classical to a quantum description.

\noindent Note 4) See \cite{SemiclIII, FileR, ACos2, SemiclIV, QuadIII, AMSS2} for the physical justification of, and mathematical methods for, 
some of the simpler simple regimes within this semiclassical scheme.

\subsection{Other Problem of Time facets within Machian Semiclassical scheme}

\noindent Configurational Relationalism {\sl remains} resolved: having reduced at the classical level, quantization does not unreduce the system.  
The classical restriction of the \K beables to a set of Dirac beables has to be abandoned. 
However, the quantum \K beables are obtained by promoting some subalgebra of the classical \K beables to quantum operators.  
And then Halliwell also provided a semiclassical construct for objects commutator-commuting with the quadratic constraint.
This is now to be used to construct a set of quantum Dirac beables as functionals of the quantum \K beables (see the next Sec).

\noindent Next, Constraint Closure remains a non-issue at the quantum level for RPM's and minisuperspace. 
Foliation and Spacetime Reconstruction issues are absent from RPM's, so we are done as regards providing {\sl a local} resolution of the PoT for this RPM's.

\subsection{Limitations of Semiclassical Approaches}

\noindent Obviously these are relatively modest through stopping short of finer/higher energy details of one's theory of Quantum Gravity.
On the other hand, semiclassical slightly inhomogeneous cosmology \cite{HallHaw} is a reasonable model for an early universe regime.  
This is a perturbative midisuperspace (inhomogeneous perturbations about minisuperspace).  
Via inflation, this might be able to explain the seeding of galaxies and CMB hot spots from quantum-cosmological fluctuations.   
Moreover, first and possibly only contact between Quantum Gravity and observational physics is likely to concern semiclassical Quantum Cosmology 
(the BICEP experiment for gravitational cosmic background radiation and eventual successors to the Planck satellite for electromagnetic cosmic background radiation).

However, even within this domain of validity, there are some problems. 
Chief among these \cite{Zeh88, Kuchar92I93} is that the working leading to such a time-dependent Schr\"{o}dinger equation 
ceases to function in the absence of making the WKB ansatz and approximation. 
This, additionally, in the quantum-cosmological context, is not known to be a particularly strongly supported ansatz and approximation to make.    
This is crucial for this Article since propping this up requires considering further PoT strategies from the classical level upwards.  
Moreover \cite{FileR} this ansatz has been shown not to hold in all regions of configuration space. 
[Though we shall concern ourselves no further with this global problem in this Article.]
The local resolution offered in the present article involves investing in Histories Theory (see Sec 9).

Other issues concern justifying the smallness of all the neglected terms.  
This should include analysis of those regimes in which one or more of these terms are not small.  
See e.g. \cite{SemiclIV} for a start on this and a list of earlier references.

\section{Summary so far of Machian strategy, with extra examples}

\noindent{\bf Level 1 (classical)}

\mbox{ }

\noindent 1) Resolve Configurational Relationalism by explicit completion of Best Matching \cite{BB82, FileR}. 
This is blocked for GR in general (Thin Sandwich Problem). 
However, it is resolved for 1- and 2-$d$ RPM's \cite{FileR}.  
It is unnecessary for minisuperspace \cite{AMSS1, AMSS2}, since here spatial homogeneity precludes nontrivial action of spatial diffeomorphisms.
It is resolved to leading order for inhomogeneous perturbations about isotropic spatially $\mathbb{S}^3$ minisuperspace with scalar field matter \cite{SIC} 
(using a Machianized version of Halliwell--Hawking's model \cite{HallHaw}).

\noindent 2) Resolving 1) then allows for one to use classical Machian emergent time $t^{\se\sm(\sJ\sB\sB)}$ classically \cite{B94I, FileR} 
to explicitly resolve the classical Frozen Formalism Problem that is induced by Temporal Relationalism. 
 
\noindent 3) The algebraic structure of the constraints then closes by good fortune for RPM's, by homogeneity for minisuperspace, and by the Dirac algebroid in the general GR case. 

\noindent 4) A second consequence \cite{AHall, FileR} of resolving 1) is that one is in possession of a set of classical \K beables. 
For minisuperspace the concept is trivial, for RPM's, these are functions of shapes, scale and their conjugate momenta, 
and see \cite{SIC} for the inhomogeneous perturbations about minisuperspace counterpart of these.
 
\noindent 5) Spacetime Relationalism is attained by the Lie derivative implementation of Diff($\FrM$) for full GR.
 
\noindent 6) Refoliation Invariance is attained by the Dirac algebroid for the full GR case.

\noindent 7) Spacetime Reconstruction is attained by exhaustion upon a family of algebraic structures, with embedding equations arising and local Lorentzian relativity emerging. 

\mbox{ } 

\noindent{\bf Level 2 (semiclassical)}

\mbox{ }  
 
\noindent 1) One hopes that classically-resolved Configurational Relationalism stays quantum-mechanically resolved (though anomalies are possible); 
this is the case for the RPM's considered, and is irrelevant in the case of minisuperspace
 
\noindent 2) The Wheeler--DeWitt equation's Frozen Formalism Problem still occurs and is not unfrozen by $t^{\se\sm(\sJ\sB\sB)}$. 
However $t^{\se\sm(\sW\sK\sB)}$ or $t^{\se\sm(\sr\se\scc)}$ can be abstracted from suitably semiclassical quantum change.

\noindent 3) Whereas we do not know how to handle quantum constraint closure in the case of general GR, for RPM's this is attained good fortune, 
whereas the minisuperspace case is aided by having only the one constraint.

\noindent 4) One either promotes one's classical level subalgebraic structure of \K beables to quantum operators or one start afresh at the quantum level.
In the case of the relational triangle case, one has (\ref{Dragt}) and conjugates as a basis of quantities for the fully quantum case, which are furtherly useful through their 
entering the kinematical quantization of the system.   
On the other hand, for minisuperspace quantum \K beables remain a trivial issue due to the absense of any linear constraints.     
 
\mbox{ }  
 
\noindent Issue 1) Justifying the WKB regime is left open at this level, as per Sec 7.2.  

\noindent Issue 2) In the absense of being able to solve the classical (or semiclassical) 1), resolutions 2) and 4) remain implicitly defined. 
We are only claiming a local resolution to apply to classical and semiclassical RPM and minisuperspace and, for now, classical-level slightly inhomogeneous cosmology.  

\noindent Issue 3) For minisuperspace \cite{AMSS1} in comoving-type coordinates privileged by the surfaces of homogeneity, 
homogeneity provides a simpler resolution of 6) and 7), both classically and quantum mechanically.
On the other hand, for RPM, 5)-7) are unnecessary since these models do not possess a GR-like notion of spacetime.  
3)-7) are nontrivially exhibited by perturbations about minisuperspace \cite{SIC}, making that a good model for these aspects, especially at quantum level for which there is not 
a known resolution for the general GR case (see also \cite{Bojo12} for recent consideration of these aspects).

\section{Level 3: Combined Machian Semiclassical--Histories--Records Approach}\label{Combo} 

Preliminarily start again with each of histories and records at the classical level, since we will be combining these with Machian classical and semiclassical approaches.  

\mbox{ }  

\noindent A) Records \cite{PW-P, EOT, Records, GMH, H99} are localized subconfigurations of a single instant that contain information/correlations. 
In a purely timeless approach, these are useful as regards reconstruction of a semblance of dynamics or history.
This is a mostly post-1993 addendum to the Introduction's Tempus Nihil Est approaches.    
  
\noindent B) Histories Theory \cite{GMH, Hartle, IL} is a path-type approach, augmented at the quantum level by attaching (projectors) projection operators to one's path. 
The decoherence functional between 2 histories is to be evaluated in terms of path integrals. 
Gell-Mann and Hartle use simple products of projectors at discrete values of label-time, whereas Isham and Linden \cite{IL} use a continuum limit of tensor products of projectors. 
The latter products succeed in themselves being a single projector. 
Thus they have the desirable feature of implementing propositions by projectors. 
This is why they are chosen for use in the combined approach. 
Isham--Linden type schemes additionally come with a classical precursor also. 
Here $\FrQ$ is supplanted by the space of histories, complete with histories momenta and histories brackets.

\mbox{ }

\noindent Then pairwise, one has I) Machian Records Theory.
\noindent II) Histories within the Machian time approach.  
\noindent III) The classical Records within Isham--Linden Histories Brackets \cite{IL} analogue of Gell-Mann--Hartle's better-known quantum inclusion of records within histories theory 
\cite{GMH, H99}. 
These two are additionally united by 

\noindent Interconnection 1): both histories and records fulfil Mackey's criterion by resting on atemporal logic \cite{IL, ID, FileR}.

\noindent Finally, the triple combination is my Machianized $\FrG$-nontrivial \cite{AHall} of Halliwell's classical prequel \cite{H03}. 
The additional interprotection at the classical level is that the classical Machian approach or histories theory `provide a semblance of dynamics or history' -- 
overcoming present-day pure records theory's principal weakness of not having well-established own means of providing such a semblance.

\noindent There is also a means of constructing classical Dirac beables (model unicorns) in extension of Halliwell's \cite{H03, AHall, FileR} as a subset amongst the quantum \K beables. 
This involves classical timeless probabilities for histories entering a region $R$ of configuration space (below; see \cite{H12} for a phase space extension). 
In the case of the relational triangle, 
\beq 
P_{R} = \int \d t^{\se\sm(\sJ\sB\sB)} \int\mathbb{D}\biP^{\mbox{\scriptsize$Dra$}} \int_{R} \mathbb{D}{\Upsilon}(\mbox{\boldmath$Dra$})\, \bn^{\mbox{\scriptsize$Dra$}} 
\cdot \biP^{\mbox{\scriptsize$Dra$}} \, w(\mbox{\boldmath$Dra$}, 
\biP^{\mbox{\scriptsize$Dra$}}) \mbox{ } , 
\eeq 
for $w$ a classical phase space distribution, $\Upsilon$ a hypersurface in configuration space with normal $\mn$. 
Then

\noindent
\beq 
A(\mbox{\boldmath$Dra$}, \mbox{\boldmath$Dra$}_0, \biP^{\mbox{\scriptsize$Dra$}}_0) := \int_{-\infty}^{+\infty} \d t^{\se\sm(\sJ\sB\sB)} \, \delta^{(3)}(\mbox{\boldmath$Dra$} - 
\mbox{\boldmath$Dra$}^{\scc\sll}(t^{\se\sm(\sJ\sB\sB)})) \mbox{ } , \mbox{ } \mbox{ } 
\eeq 
commutes with the classical constraints.


\noindent Most of the value of the combined approach, however, is at the semiclassical quantum level.  
Here the additional interprotections are as follows.

\noindent Interprotection 2) The basic idea is to prop up the principal deficiency of Level 2 -- justification of the assumption of a a WKB regime -- using decoherence in the form of 
histories decohereing \cite{Zehbook, Kieferbook}. 

\noindent Interprotection 3) As Gell-Mann and Hartle said \cite{GMH} ``{\it records are somewhere in the universe where information is stored when histories decohere}".

\noindent Interprotection 4) One can answer the elusive question of `what decoheres what' from what the records are.

\noindent Interprotection 5) By providing an underlying dynamics or history, whichever of the semiclassical Machian scheme or the histories scheme 
overcome present-day purely timeless records theory's principal weakness of needing to find a practicable construction of a semblance of dynamics or history.

\noindent Interprotection 6) The semiclassical approach provides a Machian scheme for quantum histories and quantum records to reside within. 
A prototype of this was how the Halliwell--Hawking \cite{HallHaw} semiclassical quantum cosmology scheme 
was followed up by Halliwell's work \cite{Halliwell87} on timeless correlations within such a scheme.

\noindent Interprotection 7) The semiclassical regime aids in the computation of timeless probabilities (see below for more).

\noindent At the classical level, Interprotections 2--4) are absent since they concern the purely quantum notion of decoherence, and Interprotection 7) vanishes 
since it concerns a purely quantum probability computation. 
At the quantum level, Interprotection 1) is far more significant than at the classical level too (standard logic versus Topos Theory's nontrivial intuitionistic logic \cite{ID}). 

\mbox{ }

\noindent How does the combined scheme fit together as regards primality?  
{\sl Meaningless label} histories come first; these provide the regime in which the Semiclassical Approach applies and then this in turn gives the version of the histories approach 
in which the histories run with respect to Machian semiclassical emergent time.
Then localized timeless approaches sit inside the last two of these schemes. 
On the other hand, the Semiclassical Approach sits inside the global timeless approach. 
However the global timeless approach can be taken to sit within global meaningless label time histories approach, 
so down both strands of the argument, histories are the most primary entities in the combined approach.

\mbox{ } 

\noindent Returning to Interprotection 7), this additionally provides `start afresh' means of construction of semiclassical Dirac beables (model unicorns) in extension of Halliwell's 
\cite{H03, AHall, FileR} as a subset amongst the quantum \K beables.
Here, the classical $w$ is replaced by the semiclassical quantum {\it Wigner function} $ \mbox{Wig}[\mbox{\boldmath$Dra$}, 
\biP^{\mbox{\scriptsize$Dra$}}] \approx |\chi(\mbox{\boldmath$Dra$})|^2\delta^{(3)}(\biP^{\mbox{\scriptsize$Dra$}} - \mbox{\boldmath $\pa$}S) $ by \cite{Halliwell87}, so one now has 
\beq 
P_{R}^{\sss\se\sm\si\scc\sll} \approx \int \d t^{\se\sm(\sr\se\scc)} \int_{R} \mathbb{D}\Upsilon (\mbox{\boldmath$Dra$}) \,\,\, 
\bn^{\mbox{\scriptsize$Dra$}} \cdot {\mbox{\boldmath{$\pa$}}} S \, |\chi(\mbox{\boldmath$Dra$})|^2 \mbox{ } . 
\label{36} 
\eeq 
Then 
\beq 
C_{R} := \theta
 \left( \int_{-\infty}^{\infty} \d t^{\se\sm(\sr\se\scc)} f_{R}(\mbox{\boldmath$Dra$}(t^{\se\sm(\sr\se\scc)})) - \epsilon \right) 
 P(\mbox{\boldmath$Dra$}_{\sf}, \mbox{\boldmath$Dra$}_0) \, \mbox{exp}(iS(\mbox{\boldmath$Dra$}_{\sf}, \mbox{\boldmath$Dra$}_0)) \mbox{ } , \label{gyr} 
\eeq 
commutes with the semiclassical constraints.   
This is a type of histories-theoretic {\it class functional}.\footnote{Here, cl, $0$, f superscripts denote `classical trajectory', 
initial data and final data respectively. 
$\theta$ is the step function, $f_{R}$ is the characteristic function of region $R$, $\epsilon$ is a small number, and $S$ is the classical action. 
See \cite{HT02} for the detailed form of the prefactor function $P$.} 

\mbox{ }  

\noindent See Sec 10.3 for the combined approach's own caveats and frontiers.

\section{Conclusion}\label{Conclusion} 

\subsection{Updated names for the Problem of Time facets}\label{Up-Fa} 

I identified Barbour's program as a historically-later classical precursor of part of the Problem of Time and completed that classical precursor. 
This has led to the following updated names for the Facets. 

\mbox{ }

\noindent {\bf Temporal Relationalism} is the name for the more general manifestation of the Frozen Formalism Problem.

\mbox{ }

\noindent {\bf Configurational Relationalism} is the more general name for the Thin Sandwich Problem, with Best Matching Problem being of intermediate 
generality (a Jacobi--Mach `$\mbox{\boldmath $Q$}$, d$\mbox{\boldmath $Q$}$ variables' level approach to any theory's linear constraints).

\mbox{ }

\noindent {\bf Constraint Closure Problem} is the more general name for what becomes the Functional Evolution Problem, with Partional Evolution Problem being of intermediate generality.


\noindent `{\bf Problem of Beables}' is a more cosmologically and quantum-mechanically inclusive and meaningful name and concept for the Problem of Observables.

\mbox{ }

\noindent {\bf Spacetime Relationalism} is added to the list to deal with spacetime diffeomorphisms and path integral approaches.  

\mbox{ }

\noindent The {\bf Foliation Dependence Problem} remains a fine concept and name, as does {\bf Spacetime Reconstruction Problems}, the latter acquiring plurality as per Fig 1.f).

\mbox{ }

\noindent {\bf Multiple Choice Problems}, likewise remain a fine concept and name, noting that it applies also to kinematical quantization and the 
Problem of Beables as well as to the choice of time (and frame).

\mbox{ }

\noindent {\bf Global Problems of Time}, is the final facet's renaming, emphasizing its even greater plurality: 
it can concern globality in space, time itself, spacetime, configuration space, phase space, classical solution space, Hilbert space, spaces of quantum operators...
Another classification of Global Problems of Time is into effects understandable in terms of meshing conditions of charts, of p.d.e. solutions, of representations 
or of unitary evolutions \cite{Global}.
Some forms or other of it affects almost all facets and strategies.

\subsection{A local resolution of the PoT for triangleland and some minisuperspace models} 

This article covers how a local resolution of the PoT that is, moreover, Machian in character, works out for triangleland RPM and minisuperspace models. 
In this approach, one first resolves Configurational Relationalism corresponding to the group of physically irrelevant transformations $\FrG$ acting on configurations space $\FrQ$. 
This leads to explicit expressions for the classical Machian time and for the classical \K beables. 
One then uses a Machianized version of the Semiclassical Approach to resolve frozenness at the quantum level and one promotes a subalgebra of classical \K beables to quantum ones. 
The Constraint Closure Problem, Foliation Dependence Problem and Spacetime Reconstruction problem are either absent for RPM, 
or readily overcome by use of homogeneity in the minisuperspace case. 
The WKB ansatz of the semiclassical approach is justified by histories decohereing; which degrees of freedom decohere which others is answered by looking at where the records are. 
Thus one is using a combined semiclassical-histories-records scheme such as Halliwell's \cite{H03, H09H11}. 
In fact, I use a $\FrG$-nontrivial and temporally-Machian extension of this scheme.
This scheme additionally provides separate classical and semiclassical prescriptions to form Dirac beables from one's \K beables. 
See \cite{ARel2, ACos2, SemiclIV, AHall, FileR, AMSS1, AMSS2} for further details.

\subsection{List of frontiers for the Combined Machian Approach to the Problem of Time} 

\noindent For full GR (or midisuperspace models), one is left facing the following frontiers.

\mbox{ } 

\noindent I) {\bf The Best Matching Problem becomes the Thin Sandwich Problem}. 
Thus this problem, with which there has only been some progress since Wheeler posed it in the 1960's, affects the full GR version. 
This renders the expression for t$^{\se\sm(\sJ\sB\sB)}$ and the \K beables only formal for now, and prevents use of the desired reduced, rather than Dirac, quantization.

\noindent II) {\bf Semiclassical Constraint Algebra}: looking for a parallel of the classical Dirac algebroid as regards overcoming the Constraint Closure, Foliation Dependence and 
Spacetime Reconstruction Problems.  
This is unnecessary for RPM and minisuperspace, and unknown for the general case of GR (see \cite{Bojo12} for some recent work in this direction)
It is also not clear whether the factorization of the strategizing into these three facets and `Relationalism plus beables' will continue to apply at the semiclassical level.

\noindent III) {\bf Classical and quantum combined schemes are undemonstrated from midisuperspace upward}.

\noindent IV) {\bf Likewise for the Dirac Beables constructions}.

\noindent V) One would rather use \cite{H09H11} in place of \cite{H03} in order to avoid another kind of quantum frozenness.\footnote{This is the {\it quantum Zeno problem}, 
by which `watched kettles never boil' at the quantum level \cite{Zeno}.}
However, I have not completed this yet for $\FrG$-nontrivial temporally Machian formulations.  

\noindent VI) These are for now {\it examples} of such beables, not an algebraic structure formed by these. 
Whether these cover all beables is one issue; the phase space version of the scheme is anticipated to.
What brackets these entities form remains an open question for both of Halliwell and I to look into in the future.  

\noindent VII) Halliwell-type schemes implement \cite{H03}, or part-implement \cite{H09H11}, propositions by use of configuration space (or phase space \cite{H12}) regions, 
rather than implementing these solely via projectors. 
This is problematic since classical regions do not combine in the same manner as quantum propositions \cite{FileR}.  
If it proves difficult to completely eliminate classical regions from such approaches in general, perhaps noncommutative geometry or Topos Theory \cite{ID} might help in this regard.

\noindent As regards removing the words `local' and `a' in the previous subsection's claim, 

\noindent VIII) {\bf Global Problems of Time} remain unresolved (these are {\it posed} in \cite{Global}).  

\noindent IX) {\bf Multiple Choice Problems} remain unresolved also; it is confirmed to occur for some RPM's, 
and are an unsettled question in most other models considered in this article.
%

\mbox{ }

\noindent Slightly inhomogeneous cosmology is appropriate as a next port of call for many of the above issues. 
It is not just a theoretical model, since it is sufficiently realistic to make contact with observational cosmology \cite{HallHaw}.
Here nontrivial diffeomorphism information is only considered to first order: the zeroth order (minisuperspace) needs none and the second order is discarded. 
As well as this simplification, this model arena is more tractable by the restriction to the semiclassical regime and by the splitting of this model modewise and 
into scalar, vector and tensor mode sectors.  
The current article's approach is treated for slightly inhomogeneous cosmology in \cite{SIC} 
(the classical Machian part is done, whereas the semiclassical and combined scheme parts are works in progress).  

\mbox{ } 

\noindent {\bf Acknowledgements}. I thank those close to me. 
I thank the organizers of the ``Do we need a Physics of Passage?" Conference for inviting me to give this seminar. 
I developed this work in 2012 under the support of a grant from the Foundational Questions Institute (FQXi) Fund, 
a donor-advised fund of the Silicon Valley Community Foundation on the basis of proposal FQXi-RFP3-1101 to the FQXi; Theiss Research and the CNRS administered this grant.  
John Barrow, Jeremy Butterfield, Bernard Carr, Gabriel Catren, Alexis de Saint Ours, Cecilia Flori, Tim Koslowski, Flavio Mercati and David Mulryne for further invitations, 
and them and Julian Barbour, Fay Dowker, Marc Lachi$\grave{\me}$ze-Rey, Philipp Hoehn, Chris Isham, Don Page, Brian Pitts, Carlo Rovelli, David Sloan, Rafael Sorkin, 
Reza Tavakol and various anonymous referees for discussions.
Flavio Mercati again for help with one of the Figures.


\end{document}